\documentclass{article}
\usepackage{slashed,amsmath,amssymb,enumitem}
\usepackage[papersize={8.5in,11in}]{geometry}
\begin{document}
\title{Ultraviolet Complete Electroweak Model Without a Higgs Particle}
\author{J. W. Moffat\\~\\
Perimeter Institute for Theoretical Physics, Waterloo, Ontario N2L 2Y5, Canada\\
and\\
Department of Physics and Astronomy, University of Waterloo, Waterloo,\\
Ontario N2L 3G1, Canada}
\maketitle
\begin{abstract}
An electroweak model with running coupling constants described by an energy dependent entire function is utraviolet complete and avoids unitarity
violations for energies above 1 TeV. The action contains no physical scalar fields and no Higgs particle and the physical electroweak model fields are
local and satisfy microcausality. The $W$ and $Z$ masses are compatible with a symmetry breaking $SU(2)_L\times U(1)_Y \rightarrow U(1)_{\rm em}$,
which retains a massless photon. The vertex couplings possess an energy scale $\Lambda_W > 1$ TeV predicting scattering amplitudes that can be tested
at the LHC.
\end{abstract}

%\begin{fmffile}{unifigs}

\section{Introduction}

The standard model of particle physics is very successful. However, this success is marred by the need to postulate the Higgs particle, which in the
minimal standard model is an elementary scalar spin 0 particle. The Higgs mechanism is invoked by adding to the action a complex scalar field that
transforms as an isospin doublet. A spontaneous symmetry breaking of the vacuum generates the masses of the electroweak (EW) bosons $W$ and $Z$ through a non-zero vacuum expectation value of the scalar field. The problems of the Higgs particle are well known. They include a possible triviality (or
non-interacting scalar field) problem related to the occurrence of a Landau pole for scalar fields, the existence of the hierarchy problem that causes
the Higgs mass to be unstable and the cosmological constant problem arising from the predicted vacuum energy density being many orders of magnitude
greater than the expected observational value.

Apart from these theoretical problems, a Higgs particle has not been detected experimentally. A lower bound on the Higgs mass $m_H > 114.4$ GeV has
been established by direct searches at the LEP accelerator~\cite{LEP}. The EW precision data are sensitive to $m_H$ through quantum corrections and
yield the range~\cite{EWworkgroups}:
\begin{equation}
m_H=87^{+35}_{-26}\, {\rm GeV}.
\end{equation}
The upper limit of the Higgs mass is $157$ GeV to the $95\%$ confidence-level based on using the EW data, or $186$ GeV if the LEP direct lower limit is
included. The Tevatron experiments CDF and D0 have excluded the range~\cite{Tevatron}:
\begin{equation}
158\,{\rm GeV} < m_H < 175\,{\rm GeV}.
\end{equation}
Fitting all the data, yields the result
\begin{equation}
m_H=116.4^{+15.6}_{-1.3}\, {\rm GeV},
\end{equation}
at the $68\%$ confidence level.

The question to be considered is: can we construct a physically consistent EW model containing only the observed particles, namely, 12 quarks and
leptons, the charged $W$ boson, the neutral $Z$ boson and the massless photon and gluon? Without the Higgs particle such a model is not renormalizable
and the tree graph calculation of $W_LW_L\rightarrow W_LW_L$ longitudinally polarized scattering violates unitarity above an energy of 1 TeV. We must
construct an EW model that avoids the need for it to be renormalizable and does not violate unitarity. Higgsless models have been published based on
higher-dimensional models~\cite{Grojean}, non-local regularized quantum field theory (QFT)~\cite{Moff,Moff2,MoffToth,MoffToth2} and nonlinearly
realizable group models~\cite{Ferrari}.

The motivation for introducing the Higgs particle is purely theoretical. There is no experimental evidence that this particle actually exists. Its
importance arises from its ability to generate masses of the weak $W$ and $Z$ bosons without spoiling the renormalizability of the EW gauge theory. The
LHC experiments will decide whether the simplest Higgs model is correct. If the Higgs particle is not detected, then we must consider revising at a
fundamental level the EW model of Weinberg and Salam~\cite{Weinberg,Salam,Halzen,Aitchison}. This may require a revision of our ideas about QFT.

It is possible to believe that the ``true" theory contains the renormalizable standard EW theory with a Higgs particle as a low-energy {\it effective}
theory, due to the suppression of possible non-renormalizable terms with a cut-off $\Lambda_C$. With this low-energy effective QFT interpretation, we
can use the calculational advantages of the renormalizable standard EW theory, while awaiting the detection of ``new physics" at higher energies.
However, if the Higgs particle is not detected and it is eventually decided experimentally that it does not exist, then we cannot enjoy this low-energy effective EW theory as a renormalizable theory, for it will have to be revised significantly.

In the case of renormalizable QFTs, it has been possible to remove divergences in the calculations of loop integrals by redefining the coupling
constants and masses of the particles and canceling infinities. However, there remains a large class of QFTs that are not renormalizable, such as
quantum gravity and those involving higher-dimensional operators. Attempts to regularize QFT using a procedure such as Pauli-Villars or a simple
cut-off in energy lead to a violation of gauge invariance, unitarity and Poincar\'e invariance. The dimensional regularization technique retains gauge
invariance at the cost of introducing a lower-dimensional space. This feature of strictly local QFT is partly based on the assumption that scattering
amplitudes behave no more rapidly than a polynomial as the energy increases.

We argue here that in QFT the strong assumption of polynomial behavior of amplitudes at infinity can be weakened.  This development employs the
introduction of {\it entire functions} in momentum space, which preserves unitarity, for no additional unphysical singularities are introduced at
finite energies. The amplitudes have poles or an essential singularity at infinity. The presence of an essential singularity at infinity can destroy
the process of going from Minkowski space to Euclidean space by rotating the contour of integration over the energy ($p_0\rightarrow ip_4)$. However,
regularized entire functions can be constructed that allow the QFT to be formulated from the outset in Euclidean momentum space, and then allow an
analytic continuation to Minkowski space. It is possible to formulate a relativistic, regularized QFT with entire functions that avoids all divergences in perturbation theory and maintains the unitarity and Poincar\'e invariance of the S-matrix. In these theories there is no fundamental difference
between renormalizable and non-renormalizable theories. However, there remains the question of the non-perturbative behavior of amplitudes at large
energies.

In the following, we will explore an EW model which is rendered ultraviolet (UV) finite by allowing the coupling constants $g$ and $g'$ associated with the $SU(2)_L$ and $U(1)_Y$ Feynman vertices, respectively, to possess a running energy dependence. This energy dependence is described by an entire
function ${\bar g}(p^2)=g{\cal E}(p^2/\Lambda_W^2)$, which is analytic (holomorphic) in the finite complex $p^2$ plane, and $\Lambda_W$ is an energy
scale associated with the EW interactions. This analytic property of ${\cal E}$ guarantees that no unphysical poles occur in the particle spectrum,
preserving the unitarity of the scattering amplitudes. The EW couplings $\bar e(p^2)$, $\bar g(p^2)$ and $\bar g'(p^2)$ are chosen so that off the
mass-shell ${\cal E}(p^2/\Lambda_W^2)\sim 1$ for $\Lambda_W < 1$ TeV, thereby ensuring that EW calculations at low-energies agree with experiment. On
the other hand, for energies greater than 1 TeV, ${\cal E}(p^2/\Lambda_W^2)$ decreases rapidly enough in Euclidean momentum space guaranteeing the
finiteness of radiative loop corrections. A violation of unitarity of scattering amplitudes is avoided for ${\bar g}(s)$ decreasing fast enough as a
function of the center-of mass energy $\sqrt{s}$ for $\sqrt{s} > 1$ TeV.

In a renormalizable QFT, the coupling constants run with energy as described in the renormalization group flow scenario. The energy dependence of the
coupling constants is logarithmic. In our approach, we generalize the standard renormalization group energy dependence, so that the coupling constant
energy dependence realizes a QFT finite to all orders of perturbation theory for any Lagrangian based on local quantum fields, including gravity. This
allows for a finite quantum gravity theory~\cite{QuantGrav}.

The model does not contain a fundamental scalar Higgs particle and this removes the hierarchy problem. There is no Landau pole, solving the triviality
problem for scalar fields with a quartic self-coupling. Due to the absence of a spontaneous symmetry breaking Higgs mechanism, there is no cosmological constant problem associated with a Higgs particle. Of course, there still exists a cosmological constant problem for a chiral symmetry breaking phase
in QCD with an energy scale $\Lambda_{QCD} \sim 127$ MeV, and with any other energy scale at high energy phase transitions.

\section{The Electroweak Lagrangian}

The theory introduced here is based on the local $SU(3)_c\times SU_L(2)\times U_Y(1)$ Lagrangian that includes leptons and quarks with the color degree
of freedom of the strong interaction group $SU_c(3)$. We shall use the metric convention, $\eta_{\mu\nu}={\rm diag}(+1,-1,-1,-1)$, and set $\hbar=c=1$.
The EW model Lagrangian is given by
\begin{align}
\label{Lagrangian}
{\cal L}_{\rm EM}=\sum_{\psi_L}\bar\psi_L\biggl[\gamma^\mu\biggl(i\partial_\mu - {\bar g}T^aW^a_\mu - {\bar
g}'\frac{Y}{2}B_\mu\biggr)\biggr]\psi_L\nonumber\\
+\sum_{\psi_R}\bar\psi_R\biggl[\gamma^\mu\biggl(i\partial_\mu - {\bar g}'\frac{Y}{2}B_\mu\biggr)\biggr]\psi_R
-\frac{1}{4}B^{\mu\nu}B_{\mu\nu}\nonumber\\
-\frac{1}{4}W_{\mu\nu}^aW^{a\mu\nu} + {\cal L}_M + {\cal L}_{m_f}.
\end{align}
The fermion fields (leptons and quarks) have been written as $SU_L(2)$ doublets and U(1)$_Y$ singlets, and we have suppressed the fermion generation
indices. We have $\psi_{L,R}=P_{L,R}\psi$, where $P_{L,R}=\frac{1}{2}(1\mp\gamma_5)$. Moreover,
\begin{equation}
\label{Bequation}
B_{\mu\nu}=\partial_\mu B_\nu-\partial_\nu B_\mu,
\end{equation}
and
\begin{equation}
W^a_{\mu\nu}=\partial_\mu W_\nu^a-\partial_\nu W_\mu^a-{\bar g}f^{abc}W_\mu^bW_\nu^c.
\end{equation}
The quark and lepton fields and the boson fields $W^a_\mu$ and $B_\mu$ are {\it local fields} that satisfy microcausality.

The ${\bar g}$ and ${\bar g}'$ are defined by
\begin{equation}
{\bar g}(x)=g{\cal E}(\Box(x)/\Lambda_W^2),\quad {\bar g}'(x)=g'{\cal E}(\Box(x)/\Lambda_W^2).
\end{equation}
Here, $\Lambda_W$ is an energy scale that is a measurable parameter in the model and ${\cal E}$ is an {\it entire function} of
$\Box=\partial^\mu\partial_\mu$. 

The Lagrangian for the boson mass terms is
\begin{equation}
{\cal L}_M=\frac{1}{2}M^2W^{a\mu} W^a_\mu + \frac{1}{2}M^2B^\mu B_\mu,
\end{equation}
and the fermion mass Lagrangian is
\begin{equation}
\label{fermionmass}
{\cal L}_{m_f}=-\sum_{\psi_L^i,\psi_R^j}m_{ij}^f(\bar\psi_L^i\psi_R^j + \bar\psi_R^i\psi_L^j),
\end{equation}
where $M$ and $m_{ij}^f$ denote the boson and fermion masses, respectively. Eq.(\ref{fermionmass}) can incorporate massive neutrinos and their flavor
oscillations.

Both of these mass Lagrangians explicitly break $SU(2)_L\times U(1)_Y$ gauge symmetry. In the standard model the $SU_L(2)\times U_Y(1)$ symmetry of the vacuum is spontaneously broken and the non-zero, scalar field vacuum expectation value gives mass to the fermions through an $SU_L(2)\times U_Y(1)$
invariant Yukawa Lagrangian. However, this requires 12 coupling constant parameters, which are fixed to generate the observed masses of the 12 quarks
and leptons, and a Higgs particle with a mass $m_H$ which is not predicted by the theory.

The $SU(2)$ generators satisfy the commutation relations
\begin{equation}
[T^a,T^b]=if^{abc}T^c,~~~~~\mathrm{with}~~~~~T^a=\frac{1}{2}\sigma^a.
\end{equation}
Here, $\sigma^a$ are the Pauli spin matrices and $f^{abc}=\epsilon^{abc}$. The fermion--gauge boson interaction terms are contained in
\begin{equation}
L_I=-i{\bar g}J^{a\mu}W_\mu^a-i{\bar g}'J_Y^\mu B_\mu,
\end{equation}
where the $SU(2)$ and hypercharge currents are given by
\begin{equation}
J^{a\mu}=-i\sum_{\psi_L}\bar{\psi}_L\gamma^\mu T^a\psi_L,~~~~~\mathrm{and}~~~~~J_Y^\mu=-i\sum_\psi\frac{1}{2}Y\bar\psi\gamma^\mu\psi,
\end{equation}
respectively. The last sum is over all left and right-handed fermion states with hypercharge factors $Y=2(Q-T^3)$ where $Q$ is the electric charge.
We define for later notational convenience $\slashed W=\gamma^\mu W_\mu^aT^a$.

We diagonalize the charged sector and perform mixing in the neutral boson sector. We write
$W^\pm=\frac{1}{\sqrt{2}}(W^1\mp iW^2)$ as the physical charged vector boson fields.
In the neutral sector, we can mix the fields in the usual way:
\begin{equation}
Z_\mu=\cos\theta_wW_\mu^3-\sin\theta_wB_\mu~~~\mathrm{and}~~~A_\mu=\cos\theta_wB_\mu+\sin\theta_wW_\mu^3.
\label{eq:2.35}
\end{equation}
We define the usual relations
\begin{equation}
\sin^2\theta_w=\frac{g'^2}{g^2+g'^2}~~~\mathrm{and}~~~\cos^2\theta_w=\frac{g^2}{g^2+g'^2}.
\end{equation}

If we identify the resulting $A_\mu$ field with the photon, then we have the unification condition:
\begin{equation}
e=g\sin\theta_w=g'\cos\theta_w
\end{equation}
and the electromagnetic current is
\begin{equation}
J_\mathrm{em}^\mu=J^{3\mu}+J_Y^\mu.
\end{equation}
The neutral current is given by
\begin{equation}
J_\mathrm{NC}^\mu=J^{3\mu}-\sin^2\theta_wJ_\mathrm{em}^\mu,
\end{equation}
and the fermion-boson interaction terms are given by
\begin{equation}
L_I=-\frac{\bar g}{\sqrt{2}}(J_\mu^+W^{+\mu}+J_\mu^-W^{-\mu})-{\bar g}\sin\theta_wJ_\mathrm{em}^\mu A_\mu-\frac{\bar g}{\cos\theta_w}
J_\mathrm{NC}^\mu Z_\mu.
\end{equation}

Gauge invariance is important for the QED sector, $U_{\rm em }(1)$, for it leads to a consistent quantization of QED calculations by guaranteeing that
the Ward-Takahashi identities are valid. As we will find in the later section on quantization of our EW model, quantization of the Proca massive vector boson sector of $SU(2)\times U(1)$ is physically consistent even though the $SU(2)\times U(1)$ gauge symmetry is dynamically broken~\cite{Burnel}.

\section{Symmetry Breaking}

In the standard EW model, the Higgs mechanism is chosen to make the $W^\pm$ and $Z^0$ bosons massive and the photon remains massless. To do this four
real scalar fields $\phi_i$ are introduced by adding to ${\cal L}$ an $SU(2)\times U(1)$ gauge invariant Lagrangian for the scalar fields:
\begin{equation}
\label{scalarLagrangian}
{\cal L}_\phi=\large\vert(i\partial_\mu - gT^aW^a_\mu - g'\frac{Y}{2}B_\mu)\phi\large\vert^2-V(\phi),
\end{equation}
where $\vert...\vert^2=(...)^\dagger(...)$ and
\begin{equation}
\label{potential}
V(\phi)=\mu^2\phi^\dagger\phi +\lambda(\phi^\dagger\phi)^2,
\end{equation}
where $\mu^2 < 0$ and $\lambda > 0$.
The most economical choice is to arrange the four fields $\phi_i$ in an isospin doublet of complex fields with $Y=1$. The non-zero vacuum expectation
value $v=\langle\phi\rangle_0$ has $T=\frac{1}{2}$, $T^3=-\frac{1}{2}$ and $Y=1$ and breaks $SU(2)_L\times U(1)_Y$ symmetry. The spontaneous symmetry
breaking through the Higgs mechanism generates masses for the initially massless $W$ and $Z$ bosons and provides a mechanism to generate fermion masses through a Yukawa Lagrangian.

The demand of maintaining a local gauge invariance $SU(2)\times U(1)$ as a ``hidden symmetry" through the Higgs mechanism can be viewed as a purely
aesthetic need for EW theory. However, it was recognized from the beginnings of investigations of EW models that introducing massive charged gauge
bosons $W^\pm$ in the form of mass terms $M_W^2W^{+\mu}W^-_\mu$ into the Lagrangian, produces non-renormalizable divergences. When we calculate loop
diagrams with massive bosons in standard local QFT, we get for the amplitude:
\begin{equation}
\label{loopintegral}
{\rm Amplitude}=\int d^4p({\rm propagators})\cdots.
\end{equation}
For massive boson propagators of the form:
\begin{equation}
\label{masspropagator}
iD_V^{\mu\nu}(p^2)=\frac{i\biggl(-\eta^{\mu\nu} + \frac{p^\mu p^\nu}{M^2}\biggr)}{p^2-M^2},
\end{equation}
we have for large $p^2$:
\begin{equation}
iD_V^{\mu\nu}(p^2)\sim \frac{ip^\mu p^\nu}{p^2M^2}.
\end{equation}
The integral (\ref{loopintegral}) diverges for large loop momenta by reason of the power counting of numerators and denominators in loop graphs.
Introducing a cutoff $\Lambda_C$ violates gauge invariance, Lorentz invariance and unitarity and we find that new more severe divergences in diagrams
containing more loops generate more cutoff parameters, and ultimately an infinite number of unknown parameters appears in the calculation. These
divergences cannot be renormalized and no meaningful predictions can be made in the standard local QFT.

The offending factor in the numerator of (\ref{masspropagator}) arises from the spin sum:
\begin{equation}
\label{polarization}
\sum\epsilon^\mu(p,\lambda)\epsilon^{\nu*}(p,\lambda)=-\eta^{\mu\nu}+\frac{p^\mu p^\nu}{M^2},
\end{equation}
where the polarization vector $\epsilon^\mu$ has definite spin projection $\lambda=\pm 1,0$ along the $z$-axis, while the $x$- and $y$ directions are
transverse. This corresponds to the three independent polarization vectors for a spin 1 particle. For large values of $p$ the longitudinal state
$\epsilon^\mu(p,\lambda=0)$ is proportional to $p^\mu$, leading to the numerator term $p^\mu p^\nu/M^2$ in (\ref{masspropagator}). The {\it raison
d'$\hat{e}$tre} of the spontaneous symmetry breaking Higgs mechanism is the ``gauging away" of the $p^\mu p^\nu/M^2$ term in (\ref{masspropagator}) and (\ref{polarization}), making way for a renormalizable EW theory~\cite{thooft,thooft2,thooft3,thooft4} that avoids a violation of unitarity when $\sqrt{s} > 1$ TeV. The new massive boson propagator has the form:
\begin{equation}
iD_V^{\mu\nu}(p^2)=\frac{i\biggl(-\eta^{\mu\nu}+\frac{(1-\xi)p^\mu p^\nu}{p^2-\xi M^2}\biggr)}{p^2-M^2},
\end{equation}
where $\xi$ is the gauge parameter. The dangerous factor $p^\mu p^\nu/M^2$ can now be gauged away by choosing $\xi=1$. For the ``unitary" gauge
$\xi\rightarrow\infty$, the massive spin 1 propagator reverts to (\ref{masspropagator}).

However, the spontaneous symmetry breaking mechanism can only be invoked at the price of requiring another field degree of freedom, besides the
observed $W, Z,\gamma$ bosons. This demands the existence of the Higgs particle {\it which has not been experimentally detected}. If we can construct a physically consistent QFT that is finite to all orders of perturbation theory, then we have {\it removed the primary motivation to introduce a
spontaneous symmetry breaking scenario with the related need for a renormalizable QFT}.

To circumvent predicting the existence of a Higgs particle our task is two-fold. First, we must construct a QFT that is UV complete in perturbation
theory and avoids any unitarity violation of scattering amplitudes. Secondly, we have to invoke a symmetry breaking that is {\it intrinsic to the
initial existence of $W$ and $Z$ masses and yields a massless photon}. We do not attempt to generate masses of the fermions and bosons as was done by
the spontaneous symmetry breaking of the vacuum in the standard Higgs model, or as was done in the non-local regularized EW
model~\cite{Moff,Moff2,MoffToth,MoffToth2}.

To solve the first problem, we invoke a generalized energy dependent coupling at Feynman diagram vertices:
\begin{equation}
\label{couplings}
{\bar e}(p^2)=e{\cal E}(p^2/\Lambda_W^2),\quad{\bar g}(p^2)=g{\cal E}(p^2/\Lambda_W^2),\quad {\bar g}'(p^2)=g'{\cal E}(p^2/\Lambda_W^2).
\end{equation}
Here, ${\cal E}(p^2/\Lambda_W^2)$ is an {\it entire} function for complex $p^2$ which satisfies on-shell ${\cal E}(p^2/\Lambda_W^2)=1$. This allows us
to obtain a Poincar\'e invariant, finite and unitary perturbation theory. Such entire functions are analytic (holomorphic) in the finite complex $p^2$
plane~\cite{Titchmarsh,Boas,Holland}. They must possess a pole or an essential singularity at infinity, for otherwise by Liouville's
theorem they are constant. Because they contain no poles for finite $p^2$, {\it they do not produce any unphysical particle poles and unwanted degrees
of freedom.} Provided that the vertex couplings ${\bar g}(p^2)$ and ${\bar g}'(p^2)$ decrease fast enough for $p^2\gg \Lambda_W^2$ in Euclidean
momentum space, the problem is removed of the lack of renormalizability of our minimal EW action containing only the observed twelve quarks and
leptons, the $W$ and $Z$ bosons and the massless photon.

To solve the second problem of adopting a correct economical breaking of $SU(2)\times U(1)$ symmetry, we stipulate that the massive boson Lagrangian
takes the form:
\begin{align}
\label{massmatrix}
{\cal L}_M=&\frac{1}{8}b^2g^2[(W^1_\mu)^2+(W^2_\mu)^2]+\frac{1}{8}b^2[g^2(W^3_\mu)^2-2gg'W^3_\mu B^\mu+g^{'2}B^2_\mu]\nonumber\\
=&\frac{1}{4}g^2b^2W^+_\mu W^{-\mu}+\frac{1}{8}b^2(W_{3\mu},B_\mu)\left(\begin{matrix}g^2&-gg'\\
-gg'&g^{'2}\end{matrix}\right)\left(\begin{matrix}W^{3\mu}\\B^\mu\end{matrix}\right),
\end{align}
where $b$ is the EW symmetry breaking energy scale. We see that we have the usual symmetry breaking mass matrix in which one of the eigenvalues of the
$2\times 2$ matrix in (\ref{massmatrix}) is zero, which leads to the mass values:
\begin{equation}
\label{bosonmasses}
M_W=\frac{1}{2}bg,\quad M_Z=\frac{1}{2}b(g^2+g^{\prime 2})^{1/2},\quad M_A=0.
\end{equation}
We do not identify $b$ with the vacuum expectation value $v=\langle\phi\rangle_0$ in the standard Higgs model.
The boson mass Lagrangian is given by
\begin{equation}
{\cal L}_M=M_W^2W^+_\mu W^{-\mu} + \frac{1}{2}M_Z^2Z_\mu Z^\mu.
\end{equation}

We do not know the origin of the symmetry breaking mechanism and scale $b$. To postulate the EW symmetry breaking (\ref{massmatrix}) and
(\ref{bosonmasses}) is no worse than adopting the {\it ad hoc} assumption of a scalar field Lagrangian (\ref{scalarLagrangian}) when motivating the
Higgs mechanism. There is no known fundamental motivation for choosing $\mu^2 < 0$ and we could add an additional contribution $\lambda'\phi^6$ to the
potential (\ref{potential}) or even higher order polynomials in $\phi$. The fact that such higher-dimensional operators render the EW model
non-renormalizable would not justify their lack of inclusion in our UV finite model. The quark and lepton masses and the $W$ and $Z$ masses are the
physical masses in the propagators. We avoid the problem of the lack of renormalizability of our model by damping out divergences with the coupling
vertices ${\bar g}(p^2)$, ${\bar g}'(p^2)$ and ${\bar e}(p^2)$. We emphasize that our energy scale parameter $\Lambda_W \gtrsim 1$ TeV is not a naive
cutoff. The entire function property of the coupling vertices guarantees that the model suffers no violation of unitarity
or Poincar\'e invariance.

From the relation
\begin{equation}
\label{bscale}
\frac{1}{2b^2}=\frac{g^2}{8M_W^2}=\frac{G_F}{\sqrt{2}},
\end{equation}
where $G_F=1.166\times 10^{-5}\,{\rm GeV}^{-2}$ is Fermi's constant determined from muon decay, we obtain the EW energy scale $b=246$ GeV. We now
observe that we satisfy the relation at the effective tree graph level:
\begin{equation}
\rho=\frac{M_W^2}{M_Z^2\cos^2\theta_w}=1.
\end{equation}

\section{$U(1)_{\rm em}$ Gauge Invariance and Current Conservation}

We have broken the $SU_L(2)\times U_Y(1)$ invariance of our Lagrangian to the $U_{\rm em}(1)$ invariance of QED for a massless photon. Because we have
avoided the requirement of a renormalizable local QFT for the broken $SU(2)\times U(1)$ sector, we need concern ourselves only with the need for gauge
invariance and current conservation of the QED sector $U_{\rm em}(1)$. Even though the QED sector is finite due to our regularized QED interaction, we
still demand that unphysical longitudinal modes be decoupled and the existence of Ward-Takahashi identities.

Let us consider the QED action of the form~\cite{Moff3,MoffWoodard}:
\begin{equation}
\label{QEDaction}
S_{QED}=-\int d^4x \biggl[\frac{1}{4}F^{\mu\nu}F_{\mu\nu}+\bar\psi(i\slashed\partial-m)\psi\biggr]-\int d^4xd^4y\bar\psi(x){\cal V}[eA](x,y)\psi(y),
\end{equation}
where
\begin{equation}
F_{\mu\nu}=\partial_\mu A_\nu-\partial_\nu A_\mu.
\end{equation}
The vertex operator ${\cal V}[eA]$ is in general a spinorial matrix and is formed from entire functions. It can be expanded in a power series ${\cal
V}\sim eA+(eA)^2+\cdots$. We ignore the possibility of pure photon and multifermion interactions, for they cannot be used to restore gauge invariance
and decoupling. Let us suppose that the interaction is invariant under the transformations:
\begin{equation}
\delta A_\mu(x)=-\partial_\mu\theta(x),
\end{equation}
and
\begin{equation}
\label{Toperator}
\delta\psi(x)=ie\int d^4yd^4z{\cal T}[eA](x,y,z)\theta(y)\psi(z).
\end{equation}
Here, the operator ${\cal T}\sim 1+eA+\cdots$ is a spinorial matrix and a functional of the vector potential $A_\mu$. From $\delta S_{QED}=0$, we
obtain the condition~\cite{MoffWoodard}:
\begin{align}
\frac{i}{e}\frac{\partial}{\partial y^\mu}\frac{\delta {\cal V}[eA](x,z)}{\delta A_\mu(y)}
=&-(i\slashed\partial_x-m){\cal T}(x,y,z)+\overline{\cal T}(x,y,z)(-i\overleftarrow{\slashed\partial}_z-m)\nonumber\\
+\int d^4v[{\cal V}(x,v){\cal T}(v,y,z)+\overline{\cal T}(v,y,x){\cal V}(v,z)].
\end{align}
The equations of motion are
\begin{equation}
\label{Fequation}
\partial_\nu F^{\nu\mu}(y)=\int d^4xd^4z\bar\psi(x)\frac{\delta {\cal V}[eA](x,z)}{\delta A_\mu(y)}\psi(z),
\end{equation}
and
\begin{equation}
\label{psi}
\Psi[e,A,\psi](x)\equiv (i\slashed\partial-m)\psi(x)+\int d^4z{\cal V}[eA](x,z)\psi(z)=0.
\end{equation}
Substituting (\ref{Toperator}) into the divergence of (\ref{Fequation}), we get
\begin{align}
i\partial_\mu\partial_\nu F^{\nu\mu}=\int d^4x d^4z\bar\psi(x)i\frac{\partial}{\partial y^\mu}\frac{\delta {\cal V}[eA](x,z)}{\delta
A_\mu(y)}\psi(z)\nonumber\\
=e\int d^4x d^4z[\bar\psi(x)\overline{\cal T}(z,y,x)\Psi(z)-\overline\Psi(x){\cal T}(x,y,z)\psi(z)]=0.
\end{align}
This vanishes by virtue of (\ref{psi}) and its Dirac adjoint. Current conservation establishes that a general perturbative solution exists for the QED
sector with associated Ward-Takahashi identities.

The gauge invariance and decoupling mean that a fixed covariant gauge exists such that the on-shell S matrices vanish whenever the photon polarization
vector of even one external photon is longitudinal, $\epsilon_\mu({\bf p})=-ip_\mu\theta({\bf p})$. Perturbative unitarity is also guaranteed because
of the Cutkosky rules~\cite{Cutkosky}, as long as the interactions are analytic and Hermitian. Because the photon vector transformation law is the same
as the local theory, the usual local gauge conditions are attainable. Lorentz invariance follows from having gauge fixed a manifestly invariant
theory.

A quantity that vanishes with the field equations must be proportional to them. This fact leads to the off-shell condition:
\begin{align}
\int d^4x d^4z\bar\psi(x)i\frac{\partial}{\partial y^\mu}\frac{\delta {\cal V}[eA](x,z)}{\delta A_\mu(y)}\psi(z)\nonumber\\
=e\int d^4xd^4z[\bar\psi(x)\overline{\cal W}(z,y,x)\Psi(x)-\overline\Psi(x){\cal W}(x,y,z)\psi(z)],
\end{align}
where ${\cal W}[e,A,\bar\psi,\psi]$ is a bosonic functional formed from entire functions. This suggests the invariance of the action (\ref{QEDaction})
under the transformation
\begin{align}
\label{Winvariance}
\delta A_\mu=&-\partial_\mu\theta(x),\nonumber\\
\delta\psi(x)=&ie\int d^4y d^4z{\cal W}(x,y,z)\theta(y)\psi(z).
\end{align}
So far, we have allowed for the possibility that ${\cal W}$ depends on the fermion fields. {\it However, once the action is shown to be invariant under
(\ref{Winvariance}), then we do not need the fermion field dependence for the restoration of gauge invariance or the decoupling arguments.}

Our generalized transformations form a $U(1)$ group on-shell. Our generalization of the vertex operator has not changed the group structure on-shell.
It has modified the transformation representations. The field independent representation operator of the standard local QFT:
\begin{equation}
{\cal T}[eA](x,y,z)=ie\delta^4(x-y)\delta^4(x-z),
\end{equation}
has been distorted into the field dependent form that can restore gauge invariance in our QED sector.

We have obtained a perturbatively viable, gauge invariant and finite extension of local QED. We have not found it fruitful to search for the gauge
symmetry directly. Instead, we iterate higher interactions which enforce the physical requirements of a gauge invariant QED, namely, decoupling of
unphysical modes and infer subsequently the gauge symmetry which we have shown does exist. In~\cite{MoffWoodard} the method of obtaining a generalized
gauge symmetry for QED was applied to Compton scattering. At each order of perturbation theory, decoupling is enforced on the extended Compton tree
graphs, i.e., tree amplitudes with two external fermions and $N$ external photons. This was accomplished by means of an interaction of the form:
$\bar\psi(eA)^N\psi$, which is manifestly Poincar\'e invariant, Hermitian, analytic, and sufficiently suppressed for Euclidean momentum by the vertex
operator to guarantee finiteness. The entire function ${\cal E}$ at the vertices is unity for the Nth extended Compton tree, and so it cannot affect
our ability to find an interaction.

The classical QED action is constructed to possess gauge invariance and the Ward-Takahashi identities. A problem when applying path integral
quantization can come from the measures: $[dA], [d\psi]$ and $[d\bar\psi]$. The ordinary local photon transformation remains unchanged, $[dA]$ is
invariant and gauge fixing can be done as in local QFT. The Grassmann variables give the transformation rule to lowest order in
$\theta$~\cite{MoffWoodard}. We define a dot product:
\begin{equation}
(\theta\cdot{\cal T}[eA])(x,z)=\int d^4y\theta(y){\cal T}[eA](x,y,z).
\end{equation}
\begin{equation}
[d\psi']=[d\psi]{\rm det}^{-1}(1+ie\theta\cdot{\cal T}[eA])=[d\psi]\exp[-ie{\rm Tr}(\theta\cdot{\cal T}[eA])],
\end{equation}
where the trace involves summing over spinor indices and integrating over spacetime coordinates. We have
\begin{equation}
[d\psi'][d\bar\psi']=[d\psi][d\bar\psi]\exp[-ie{\rm Tr}(\theta\cdot{\cal T}[eA])+ie{\rm Tr}(\theta\cdot\overline{\cal T}[eA])].
\end{equation}
The non-invariance that arises from this result is absorbed by a measure factor $\mu[eA]$:
\begin{equation}
\mu[eA]\equiv \exp(iS_{\rm meas}[eA]),
\end{equation}
where we require that
\begin{equation}
\partial_\mu\frac{\delta S_{\rm meas}[eA]}{\delta A_\mu}=-e{\rm Tr}(\theta\cdot{\cal T}[eA]+{\rm Tr}(\theta\cdot\overline{\cal T}[eA])).
\end{equation}

\section{Quantization of the EW Model}

We do not attempt to explain the origin of the $W$ and $Z$ boson masses through e.g., a Higgs field spontaneous symmetry breaking mechanism. Instead, we treat the masses of the $W$ and $Z$ bosons as intrinsic to our EW model by assuming a Proca action~\cite{Greiner}. For the spin 1 boson fields $W$ and $Z$ this requires that we isolate the relevant degrees of freedom. We have the canonically conjugate fields:
\begin{equation}
\pi_B^\mu=\frac{\partial{\cal L}}{\partial(\partial_0B_\mu)}=-B^{0\mu},\quad \pi_W^{a\mu}=\frac{\partial{\cal
L}}{\partial(\partial_0W^a_\mu)}=-W^{a0\mu},
\end{equation}
or
\begin{equation}
\pi_B^0=0,\quad \pi_B^i=-B^{0i},\quad {\rm and}\quad \pi_W^{a0}=0,\quad \pi_W^{ai}=-W^{a0i}\quad (i=1,2,3).
\end{equation}

The Proca fields have only three independent dynamical degrees of freedom. This can be seen from the equation of motion for the $B_\mu$ field:
\begin{equation}
\partial_\mu B^{\mu\nu}+M^2B^\nu=J_Y^\nu,
\end{equation}
which can be written as
\begin{equation}
\Box B^\nu-\partial^\nu(\partial_\mu B^\mu)+M^2 B^\nu=J_Y^\nu.
\end{equation}
The four-divergence of this equation gives
\begin{equation}
\partial_\nu B^\nu=\frac{1}{M^2}\partial_\nu J^\nu_Y.
\end{equation}
The source current $J^\nu_Y$ need not be conserved for the Proca field. However, if we assume that it is, $\partial_\nu J^\nu_Y=0$, then we have that
\begin{equation}
\label{Lorenz}
\partial_\nu B^\nu=0,
\end{equation}
is automatically satisfied. The Lorenz condition becomes a constraint equation for the Proca field, making the $B^0$ a dependent variable. We have the
Proca equation for $J_Y^\nu=0$:
\begin{equation}
\partial_\mu B^{\mu0}+M^2B^0=0.
\end{equation}
This yields the equation
\begin{equation}
\label{Bconstraint}
B^0=-\frac{1}{M^2}\partial_iB^{i0},
\end{equation}
which shows that $B^0$ is a dependent quantity and not an independent dynamical degree of freedom.
The Hamiltonian for the $B^\mu$ field is given by
\begin{equation}
H_B=\int d^3x\frac{1}{2}\biggl[(B^{i0})^2+(B^{ij})^2+M^2(B^{i0})^2+\frac{1}{M^2}(\partial_iB^{i0})^2\biggr].
\end{equation}

Let us now turn to the non-Abelian gauge field $W^a_\mu$. The covariant derivative operator is given by
\begin{equation}
D^\mu W^a_{\mu\nu}\equiv \partial^\mu W^a_{\mu\nu}+igf^{abc}W^{b\mu}W^c_{\mu\nu}.
\end{equation}
The equations of motion are
\begin{equation}
D_\mu W^{a\mu\nu}+M^2W^{a\nu}=J^{a\nu}.
\end{equation}
Taking $J^{a\nu}=0$ and $\nu=0$ gives
\begin{equation}
\label{Wconstraint}
W^{a0}=-\frac{1}{M^2}D_iW^{ai0}.
\end{equation}
As with the $U(1)$ Abelian field $B^\mu$ the $W^{a0}$ is not an independent dynamical degree of freedom. The Hamiltonian for the $W^{a\mu}$ field is
\begin{equation}
H_W=\int d^3x\frac{1}{2}\biggl[(W^{ai0})^2+(W^{aij})^2+M^2(W^{ai0})^2+\frac{1}{M^2}(D_iW^{ai0})^2\biggr].
\end{equation}

A covariant quantization of the Proca fields can be derived by imposing the second class constraints on the field operators using (\ref{Bconstraint})
and (\ref{Wconstraint}) as operator constraints~\cite{Dirac}. We have the equal time commutation relations:
\begin{align}
[B^{i0}({\bf x},t),B^0({\bf x}',t))&=\frac{i}{M^2}{\bf\nabla}^i\delta^3({\bf x}-{\bf x}'),\nonumber\\
[B^0({\bf x},t),B^0({\bf x}',t)]&=0,
\end{align}
and
\begin{align}
[W^{ai0}({\bf x},t),W^{b0}({\bf x}',t)]&=\delta^{ab}\frac{i}{M^2}{\bf\nabla}^i\delta^3({\bf x}-{\bf x}'),\nonumber\\
[W^{a0}({\bf x},t),W^{b0}({\bf x}',t)]&=0.
\end{align}
We now obtain the covariant commutation relation for the $B^\mu$ field:
\begin{equation}
[B^\mu(x),B^\nu(y)]=-i\biggl(\eta^{\mu\nu}+\frac{1}{M^2}\partial^\mu\partial^\nu\biggr)D(x-y),
\end{equation}
where $D(x-y)$ is the Pauli-Jordan propagator:
\begin{equation}
iD(x-y)=\int\frac{d^4p}{(2\pi)^3}\epsilon(p_0)\delta(p^2-M^2)\exp[-ip\cdot (x-y)].
\end{equation}
For the $W^a_\mu$ field we have
\begin{equation}
[W^{a\mu}(x),W^{b\nu}(y)]=-i\delta^{ab}\biggl(\eta^{\mu\nu}+\frac{1}{M^2}\partial^\mu\partial^\nu\biggr)D(x-y).
\end{equation}

The fermion propagator of our EW theory in momentum space is given by
\begin{equation}
\label{fermionprop}
iS=\frac{-i}{\slashed p-m+i\epsilon}.
\end{equation}
For massless photons we have
\begin{equation}
\label{photonprop}
iD_{\gamma\mu\nu}=\frac{i}{p^2+i\epsilon}\biggl(-\eta_{\mu\nu}+(1-\xi)\frac{p_\mu p_\nu}{p^2}\biggr),
\end{equation}
while for the massive $W$ and $Z$ bosons:
\begin{equation}
\label{vectorprop}
iD_{V\mu\nu}=\frac{i}{p^2-M^2+i\epsilon}\biggl(-\eta_{\mu\nu}+\frac{p_\mu p_\nu}{M^2}\biggr).
\end{equation}

The quantization of our EW model can be achieved using a path integral formalism with a measure that can be chosen to maintain the gauge invariance of
the model for the QED massless case. A measure can be chosen that is consistent with the intrinsic dynamical symmetry breaking $SU_L(2)\times
U_Y(1)\rightarrow U_{\rm em}(1)$. This will be a topic of further research in a future paper.

\section{Properties of the Entire Function ${\cal E}$}

The standard physical requirements of relativistic field theory are obtained if we choose tempered test functions. For a field $\phi(x)$ an
operator-valued generalized function, averaged over a smooth test function $f(x)$ satisfies:
\begin{equation}
\phi=\int dx{\tilde \phi}(x)f(x).
\end{equation}
This property of test functions reflects the symmetry between coordinate and momentum space. The temperedeness of functions leads to scattering
amplitudes being analytic in $s$ ($s=({\rm center-of-mass\,\,energy})^2$) for fixed $t < 0$ ($t=({\rm momentum-transfer})^2$) in a cut plane and it
possesses polynomial behavior. These are conditions satisfied by {\it strictly} local quantum field theory.

A local field theory in the framework of axiomatic field theory, satisfies the following requirements:
\begin{enumerate}[label=\Roman{*}/, ref=(\roman{*})]

\item  A Hilbert space of states,

\item  The fields are invariant under the Poincar\'e group of transformations,

\item  The fields satisfy local commutativity,

\item  Positive energy,

\item  A particle interpretation,

\item  The scattering amplitudes satisfy unitarity.

\end{enumerate}

We can use non-tempered test functions consistent with (I) - (VI)~\cite{Jaffe,Jaffe2}. The scattering amplitudes for these functions can grow, for large energies, faster than any polynomial. Such functions are described by {\it entire functions}~\cite{Titchmarsh,Boas,Holland}. Strictly localizable fields demand test functions with compact support in configuration space. This requires test functions in momentum space, which decrease at infinity as
$\exp(-||p ||^a)$ with $a < 1$, where $|| p ||$ is the Euclidean norm.

Local commutativity can be widened to include values $a\leq 1$. Wightman functions can grow arbitrarily fast near the light cone for fields that are
not strictly localizable, and still satisfy a condition of microcausality~\cite{Meiman,Efimov,Hieu,Taylor,Constantinescu,Taylor2}.

Consider operators ${\cal E}(t)$ represented as infinite series in powers of $t=-p^2$:
\begin{equation}
{\cal E}(t)=\sum_{j=0}^\infty\biggl[\frac{|c_j^2|}{(2j)!}\biggr]^{1/2}t^j.
\end{equation}
The function ${\cal E}(t)$ is an {\it entire} function of $t$, which is analytic (holomorphic) in the finite complex $t$ plane. Thus, ${\cal E}$ has no singularities in the finite complex $t$ plane. However, it will have a pole or an essential singularity at infinity, otherwise, by
Liouville's theorem it is constant. This avoids non-physical singularities occurring in scattering amplitudes which violate the S-matrix unitarity and
the Cutkosky rules~\cite{Cutkosky}.

For the entire functions ${\cal E}(t)$, we can distinguish three cases:
\begin{enumerate}
\item  \quad\quad ${\rm lim\,sup}_{j\rightarrow\infty}\vert c_j\vert^{\frac{1}{j}}=0,$
\item  \quad\quad ${\rm lim\,sup}_{j\rightarrow\infty}\vert c_j\vert^{\frac{1}{j}}={\rm const. }< \infty,$
\item  \quad\quad ${\rm lim\,sup}_{j\rightarrow\infty}\vert c_j\vert^{\frac{1}{j}}=\infty.$
\end{enumerate}
Let us consider the order of the entire functions ${\cal E}(t)$. For (1) the order is $\gamma < \frac{1}{2}$:
\begin{equation}
\vert{\cal E}(t)\vert < \exp(\alpha\vert t\vert^\gamma),
\end{equation}
where $\alpha > 0$. An entire function with this property is
\begin{equation}
{\cal E}(z)=\sum^\infty_{n=0}\frac{z^n}{\Gamma(bn+1)}\quad b > 2,
\end{equation}
where $\Gamma$ is the gamma function. For type (1) entire functions, we know that ${\cal E}(t)$ does not decrease along any direction in the complex
$z$ plane. Therefore, we cannot use this type of function to describe our coupling functions ${\bar e}(t)$, ${\bar g}(t)$ and ${\bar g}'(t)$, for they
will not produce a UV finite perturbation theory.

For case (2), the entire functions ${\cal E}(t)$ are of order $\gamma=1/2$ and we have
\begin{equation}
\vert{\cal E}(t)\vert \leq \exp(\alpha\sqrt{|t|}).
\end{equation}
This type of entire function {\it can decrease along one direction} in the complex $t$ plane.

In the case (3), we have $\gamma > 1/2$ and now
\begin{equation}
\vert{\cal E}(t)\vert \leq \exp(f(t)\vert t\vert),
\end{equation}
where $f(\vert t\vert)$ is a positive function which obeys the condition $f(\vert t\vert) > \alpha\vert t\vert^{1/2}$ as $\vert t\vert\rightarrow
\infty$ for any $\alpha > 0$. These functions can decrease along whole regions for $\vert t\vert\rightarrow \infty$ and can be chosen to describe the
coupling functions and lead to a UV finite, unitary perturbation theory. We note that a consequence of the fundamental theorem of algebra is that
``genuinely different" entire functions {\it cannot dominate each other}, i.e., if $f$ and $g$ are entire functions and $|f| \leq |g|$ everywhere, then $f=ag$ for some complex number $a$. This theorem plays an important role in providing a uniqueness of choice of entire functions ${\cal
E}(p^2/\Lambda_W^2)$ for $p^2\rightarrow\infty$.

Let us consider operators ${\cal E}$ of the type~\cite{Efimov,Efimov2,Efimov3}:
\begin{equation}
{\cal E}(x-y)={\cal E}(\Box(x))\delta^4(x-y),
\end{equation}
where ${\cal E}(\Box)$ has the integral representation:
\begin{equation}
{\cal E}(\Box)=\int_{r^2<\Lambda^2}d^4r\kappa(r^2)\exp\biggl[ir_0\frac{\partial}{\partial{x^0}}
+ {\bf r}\cdot\frac{\partial}{\partial{\bf x}}\biggr]
$$ $$
=(2\pi)^2\int^\Lambda_0d\beta\beta^2\kappa(\beta^2)\frac{J_1(\beta(\Box^{1/2}))}{\Box^{1/2}}.
\end{equation}
We can also have the integral representation:
\begin{equation}
{\cal E}(\Box)=\int_{r^2<\Lambda^2}d^4r\kappa(r^2)\exp\biggl[r_0\frac{\partial}{\partial{x^0}} + i{\bf r}\cdot\frac{\partial}{\partial{\bf x}}\biggr]
$$ $$
=(2\pi)^2\int^\Lambda_0d\beta\beta^2\kappa(\beta^2)\frac{J_1(\beta(-\Box^{1/2}))}{(-\Box^{1/2})}.
\end{equation}
The function $\kappa(r^2)$ is an integrable function of the Euclidean 4-vector $r$ with $r^2=r_0^2+r_1^2+r_2^2+r_3^2$ and $J_1(z)$ is a Bessel
function. Moreover, $\Lambda$ denotes a fundamental length in the QFT. In the momentum representation we have
\begin{equation}
\label{AA}
{\cal E}(t)=(2\pi)^2\int^\Lambda_0d\beta\beta^2\kappa(\beta^2)\frac{J_1(\beta(t)^{1/2})}{(t)^{1/2}}.
\end{equation}
We also have
\begin{equation}
\label{BB}
{\cal E}(t)=(2\pi)^2\int^\Lambda_0d\beta\beta^2\kappa(\beta^2)\frac{J_1(\beta(-t)^{1/2})}{(-t)^{1/2}}.
\end{equation}
For operators of type (\ref{AA}), the ${\cal E}(t)$ decrease as $t\rightarrow \infty$ and increase as $t\rightarrow -\infty$, while for (\ref{BB}) they
decrease as $t\rightarrow -\infty$ and increase as $t\rightarrow \infty$.

We introduce a regularizing function $R^\delta(t)$ which approximates distributions~\cite{Efimov}:
\begin{equation}
\label{delta}
{\cal E}^\delta(x)=\frac{1}{(2\pi)^4}\int d^4p\exp(ip\cdot x){\cal E}(t)R^\delta(t).
\end{equation}
Here,
\begin{equation}
R^\delta(t)=\exp[-\delta(t+iN^2)^{1/2+\nu}\exp(-i\pi\sigma)],
\end{equation}
where
$0 < \nu < \sigma < 1/2$ and $N$ is a positive parameter. For large $t$ we obtain
\begin{equation}
|R^\delta(t)|\sim \exp\{-\delta|t|^{1/2+\nu}\cos[\pi\sigma-(\nu+\frac{1}{2}){\rm arg}t]\}.
\end{equation}
We see that $R^\delta$ is an analytic function that falls off faster than a linear exponential in the upper-half plane of the complex variable $t$. The
integral in (\ref{delta}) is convergent for $\delta > 0$, so that ${\cal E}^\delta(x-y)$ is well behaved. By using the regularizing functions
$R^\delta$, we can perform a rotation over $p_0$ by an angle $\pi/2$ in the integral:
\begin{equation}
(G^\delta,f)=-i\int d^4p\exp(ip\cdot x){\tilde f}(p)\int d^4q{\cal E}^{(1)\delta}(-q^2){\cal E}^{(2)\delta}(-(p-q)^2),
\end{equation}
where
\begin{equation}
{\cal E}^\delta(-q^2)={\cal E}(-q^2)R^\delta(q^2),
\end{equation}
and the function $G^\delta$ is such that in the limit $\delta\rightarrow 0$, the functional $(G^\delta,f)$ is well defined for test functions $f$.

The coupling functions at the vertices of Feynman graphs will in momentum space have the behavior ${\bar g}(p^2)=g{\cal E}(p^2)$ and we require for the
convergence of loop integrals that in Euclidean momentum space for $p^2\rightarrow\infty$ the coupling ${\bar g}(p^2)$ vanishes fast enough to
guarantee convergence of the integrals. We emphasize that the coupling functions ${\bar g}(p^2)$ and ${\bar g'}(p^2)$ {\it are not the Fourier transforms of physical field operators in the
action}.

\section{Running Coupling Constants}

The Feynman rules for EW interactions will make use of the propagators (\ref{fermionprop}), (\ref{photonprop}) and (\ref{vectorprop}) and the vertex
factor for electromagnetic interactions:
\begin{equation}
-ieQ_f\gamma^\mu\rightarrow -i{\bar e}(p^2)Q_f\gamma^\mu,
\end{equation}
where $Q_f$ is the charge of the fermion $f$: $Q_f=-1$ for the electron. The outgoing $\bar f$ should be drawn as an ingoing $f$ with $\bar u(p)$ and
$u(p)$ spinors attached to the fermion lines.  For the charged current interactions we have for virtual particle exchanges the vertex factor
replacement:
\begin{equation}
-i\frac{g}{\sqrt{2}}(\bar\psi_L\gamma^\mu T^\pm\psi_L)\rightarrow -i\frac{{\bar g}(p^2)}{\sqrt{2}}(\bar\psi_L\gamma^\mu T^\pm\psi_L).
\end{equation}
For the neutral current interaction, we have
\begin{equation}
-i\frac{g}{\cos\theta_w}\gamma^\mu\frac{1}{2}(C^f_v-C^f_a\gamma^5)\rightarrow -i\frac{{\bar
g}(p^2)}{\cos\theta_w}\gamma^\mu\frac{1}{2}(C^f_v-C^f_a\gamma^5).
\end{equation}
In the above, ${\bar e}(p^2)$, ${\bar g}(p^2)$ and ${\bar g}'(p^2)$ are given by (\ref{couplings}). They satisfy at low energies for $p^2 <
\Lambda_W^2$ with $\Lambda_W > 1$ TeV:
\begin{equation}
{\bar e}(p^2)\sim e,\quad {\bar g}(p^2)\sim g,\quad {\bar g}'(p^2)\sim g',
\end{equation}
which assures that all low-energy EW calculations at the tree diagram level agree with EW data. Thus,
tree graph decay processes such as $W^-\rightarrow e^- +{\bar\nu}_e$ are calculated as in the SM using the vertex:
\begin{equation}
-\frac{ig}{\sqrt{2}}\gamma^\mu\frac{(1-\gamma_5)}{2},
\end{equation}
with the predicted $W$ width:
\begin{equation}
\Gamma(W^-\rightarrow e^-{\bar\nu}_e)=\frac{1}{12}\frac{g^2}{4\pi}M_W\sim 205\,{MeV}.
\end{equation}

Let us consider the scattering of electrons by a static charge~\cite{Halzen}. The covariant amplitude is given by
\begin{equation}
\label{Rutherford}
-i{\cal M}=(i{\bar e}(p^2){\bar u}\gamma^\mu u)\biggl(\frac{-i\eta_{\mu\nu}}{p^2}\biggr)(-ij^\nu({\bf p})),
\end{equation}
where for a static nucleus of charge $Ze$: $j^0({\bf x})=\rho_c({\bf x})=Ze\delta({\bf x})$ and $p^2=-|{\bf p}|^2$. The photon-electron vertex coupling
${\bar e}(p^2)$ is given by (\ref{couplings}) and the energy scale $\Lambda_W$ for electromagnetic processes is $\Lambda_W > 1$ TeV.

By including a one-loop photon correction we obtain
\begin{align}
-i{\cal M}=-(ie{\cal E}(p^2/\Lambda^2_W){\bar u}\gamma^\mu u)\biggl(\frac{-i\eta_{\mu\mu'}}{p^2}\biggr)
\int \frac{d^4k}{(2\pi)^4}\biggl[(ie{\cal E}(k^2/\Lambda^2_W)\gamma^{\mu'}_{\alpha\beta}\nonumber\\
\times\frac{i(\slashed k+m)_{\beta\lambda}}{k^2-m^2}(ie{\cal E}((k-p)^2/\Lambda_W^2))\gamma^{\nu'}_{\lambda\tau}\frac{i(\slashed k-\slashed p +
m_e)_{\tau\alpha}}{(k-p)^2-m_e^2}\biggr]\biggl(\frac{-i\eta_{\nu'\nu}}{p^2}\biggr)(-ij^\nu({\bf p})).
\label{oneloop}
\end{align}
By adding (\ref{oneloop}) to (\ref{Rutherford}), we have modified the propagator:
\begin{equation}
\frac{-i\eta_{\mu\nu}}{p^2}\rightarrow \frac{-i\eta_{\mu\nu}}{p^2}
+ \biggl(\frac{-i\eta_{\mu\mu'}}{p^2}\biggr)\Pi^{\mu'\nu'}_{\gamma\gamma}(p^2)\biggl(\frac{-i\eta_{\nu'\nu}}{p^2}\biggr),
\end{equation}
where
\begin{align}
\Pi_{\gamma\gamma\mu\nu}(p^2)=-\int \frac{d^4k}{(2\pi)^4}{\rm Tr}\biggl\{(ie{\cal E}(k^2/\Lambda_W^2)\gamma_\mu)\frac{i(\slashed k +
m_e)}{k^2-m_e^2}(ie{\cal E}((k-p)^2/\Lambda_W^2)\gamma_\nu)\nonumber\\
\times\frac{i(\slashed k-\slashed p+m_e)}{(k-p)^2-m_e^2}\biggr\}.
\label{vacpol}
\end{align}
We assume that the entire function ${\cal E}(k^2/\Lambda_W^2)$ decreases in Euclidean momentum space fast enough for $k^2\gg \Lambda_W^2$ to make the
integral over $k$ in (\ref{vacpol}) converge.

The vacuum polarization tensor is defined by
\begin{equation}
\Pi^{\mu\nu}(p^2)=\Pi^T(p^2)\biggl(\eta^{\mu\nu}-\frac{p^\mu p^\nu}{p^2}\biggr)+\Pi^L(p^2)\frac{p^\mu p^\nu}{p^2},
\end{equation}
where $\Pi^T(p^2)$ and $\Pi^L(p^2)$ are the transverse and longitudinal parts of $\Pi^{\mu\nu}(p^2)$, respectively.
A calculation yields
\begin{equation}
\label{PiTransverse}
\Pi^T_{\gamma\gamma\mu\nu}(p^2)=-\eta_{\mu\nu}p^2\Pi_{\gamma\gamma}^T(p^2),
\end{equation}
with
\begin{equation}
\label{TransVac}
\Pi^T_{\gamma\gamma}(p^2)=\frac{2\alpha}{\pi}\int^1_0dxx(1-x)F\biggl(x,\frac{p^2}{\Lambda_W^2}\biggr),
\end{equation}
where $\alpha=e^2/4\pi$. We note from (\ref{PiTransverse}) that $\Pi^T_{\gamma\gamma\mu\nu}(0)=0$ in accordance with the Ward-Takahashi identity valid
for the $U_{\rm em}(1)$ gauge invariance. The gauge invariance to $O(e^2)$ is guaranteed by choosing a suitable vertex correction diagram that restores
invariance to this order~\cite{MoffWoodard}.

For $p^2\ll \Lambda_W^2$, we obtain from (\ref{TransVac}):
\begin{equation}
\Pi^T_{\gamma\gamma}(p^2)\simeq \frac{\alpha}{3\pi}\ln\biggl(\frac{\Lambda_W^2}{m_e^2}\biggr)+\frac{\alpha}{15\pi}\frac{p^2}{m_e^2}.
\end{equation}
For large $p^2$, we get
\begin{equation}
\label{largeF}
\frac{2\alpha}{\pi}\int^1_0dxx(1-x)F\biggl(x,\frac{p^2}{\Lambda_W^2}\biggr)\simeq I(p^2).
\end{equation}

For Rutherford scattering, including the loop correction, the amplitude is given for small $p^2$ by
\begin{equation}
-i{\cal M}=(ie\bar u\gamma_0u)\biggl(\frac{-i}{p^2}\biggr)\biggl[1-\frac{\alpha}{3\pi}\ln\biggl(\frac{\Lambda_W^2}{m_e^2}\biggr)
-\frac{\alpha}{15\pi}\frac{p^2}{m_e^2}+O(e^4)\biggr](-iZe).
\end{equation}
We can now perform for small $p^2\ll\Lambda_W^2$ a {\it finite} renormalization:
\begin{equation}
-i{\cal M}=(ie_R\bar u\gamma_0u)\biggl(\frac{-i}{p^2}\biggr)\biggl(1-\frac{e_R^2}{60\pi^2}\frac{p^2}{m_e^2}\biggr)(-iZe_R),
\end{equation}
where
\begin{equation}
e_R\equiv e\biggl[1-\frac{e^2}{12\pi^2}\ln\biggl(\frac{\Lambda_W^2}{m_e^2}\biggr)\biggr]^{1/2}.
\end{equation}
Here, $e_R$ is the renormalized charge and the measured fine-structure constant is: $\alpha_R=e_R^2/4\pi$.

The interaction between the electron and the renormalized charge $Ze_R$ is described by the potential~\cite{Halzen}:
\begin{equation}
V(r)=-\frac{Ze_R^2}{4\pi r}-\frac{Ze_R^4}{60\pi^2m_e^2}\delta(\bf r).
\end{equation}
The screening of the charged nucleus leads to the Lamb shift:
\begin{equation}
\Delta E_{nl}=-\frac{e_R^4}{60\pi^2m_e^2}|\psi_{nl}(0)|^2=-\frac{8\alpha^3_R}{15\pi n^3}Ry\delta_{l0},
\end{equation}
where the $\psi_{nl}$ are the hydrogen atomic wavefunctions and $Ry=m_e\alpha_R^2/2$ is the Rydberg constant. This result together with additional loop
corrections calculated for small $p^2$ reproduces the accurately measured Lamb shift.

Our modified dependence of $I(p^2)$ in (\ref{largeF}) at large $p^2$ will differ from the standard result obtained for the $\Pi^T_{\gamma\gamma}(p^2)$
calculated for ${\cal E}(p^2/\Lambda_W^2)=1$. The usual QED result is given by
\begin{equation}
I(p^2)\simeq \frac{\alpha}{3\pi}\biggl[\ln\biggl(\frac{\Lambda_C^2}{m_e^2}\biggr)-\ln\biggl(\frac{p^2}{m_e^2}\biggr)\biggr]=
\frac{\alpha}{3\pi}\ln\biggl(\frac{\Lambda_C^2}{p^2}\biggr),
\end{equation}
where $\Lambda_C$ is a cutoff. However, for us the contribution $(\alpha/3\pi)\ln(\Lambda_W^2/m_e^2)$ is finite and is absorbed by the charge by a
finite renormalization, whereas in the standard QED the contribution $(\alpha/3\pi)\ln(\Lambda_C^2/m_e^2)$ is infinite as $\Lambda_C\rightarrow\infty$,
and results in an infinite renormalization of the charge $e$.

In our perturbation theory, the relation between $e^2(p^2)$ and the bare charge $e_0^2$ is determined by
\begin{equation}
e^2(p^2)=e_0^2[1-\Pi^T_{\gamma\gamma}(p^2) + O(e_0^4)],
\end{equation}
where for us $\Pi^T_{\gamma\gamma}(p^2)$ is a finite quantity and the physical energy scale $\Lambda_W$ will be determined by measurement. We can
obtain the running coupling constant $e(p^2)$ in terms of the ``bare" charge $e_0$ by summing all orders of perturbation theory:
\begin{equation}
e^2(p^2)=e_0^2\biggl(\frac{1}{1+\Pi^T_{\gamma\gamma}(p^2)}\biggr).
\end{equation}
We now have for large $p^2$ from (\ref{largeF}):
\begin{equation}
\label{runningconst}
\alpha(p^2)=\frac{\alpha_0}{1+I(p^2)},
\end{equation}
where $\alpha_0=e_0^2/4\pi$ is the bare fine-structure constant. Depending on the sign of $I(p^2)$ a Landau pole can occur in $\alpha(p^2)$. For the
usual QED case, we have for a renormalization group energy scale $\mu$:
\begin{equation}
I(p^2)\simeq -\frac{\alpha(\mu^2)}{3\pi}\ln\biggl(\frac{p^2}{\mu^2}\biggr)
\end{equation}
and a Landau pole occurs for $p^2\simeq\mu^2\exp(3\pi/\alpha(\mu^2))$.

Let us calculate the photon vacuum polarization tensor $\Pi_{\gamma\gamma}^{\mu\nu}$, assuming that the virtual fermion loop is dominated by the top
quark. We choose as the entire function in Euclidean momentum space:
\begin{equation}
\label{EntireFunction}
{\cal E}(p^2/\Lambda_W^2)=\exp\biggl(-\frac{p^2+m_t^2}{\Lambda_W^2}\biggr),
\end{equation}
where $m_t$ denotes the top quark mass: $m_t=173.1\pm 1.3$ GeV~\cite{pdg}. A calculation yields~\cite{MoffWoodard}:
\begin{equation}
\label{gammavac}
\Pi^T_{\gamma\gamma}(p^2)=\frac{4\alpha}{\pi}\int_0^{1/2} dxx(1-x)E_1\biggl(x\frac{p^2}{\Lambda_W^2} + \frac{1}{1-x}\frac{m_t^2}{\Lambda_W^2}\biggr),
\end{equation}
where $E_1$ is the exponential integral:
\begin{equation}
E_1(z)\equiv \int^\infty_z dt\frac{\exp(-t)}{t}=-\ln(z)-\gamma-\sum^\infty_{n=1}\frac{(-z)^n}{nn!},
\end{equation}
and $\gamma$ is the Euler-Masheroni constant. We note that, as before, the factor of $p^2$ in (\ref{PiTransverse}) guarantees the masslessness of the
photon and it satisfies the QED Ward-Takahashi identity: $\Pi^T_{\gamma\gamma\mu\nu}(0)=0$. Also, the $U(1)_{\rm em}$ gauge invariance has absorbed the
naive quadratic divergence.

We now develop an asymptotic expansion of (\ref{gammavac}) for $p^2\ll \Lambda_W^2$:
\begin{align}
\label{regularized}
\Pi^T_{\gamma\gamma}(p^2)=\frac{e^2}{2\pi^2}\biggl[\frac{1}{6}\ln\biggl(\frac{\Lambda_W^2}{p^2}\biggr)+\frac{1}{6}\ln(2\pi)-\frac{1}{6}
\gamma -\frac{13}{72}\nonumber\\
-\int^1_0 dxx(1-x)\ln\biggl[x(1-x)p^2+m_t^2\biggr] + O\biggl(\frac{\ln(\Lambda^2_W)}{\Lambda_W^2}\biggr)\biggr].
\end{align}
We can compare this result with the same calculation performed using dimensional regularization in $D$ dimensions with energy scale $\mu$ in standard
local QED with ${\cal E}(p^2/\Lambda_W^2)=1$:
\begin{align}
\Pi^T_{\gamma\gamma}(p^2)=e^22^{D/2+1}\frac{\Gamma(2-D/2)}{2^D\pi^{D/2}}\int^1_0 dxx(1-x)\biggl[x(1-x)\frac{p^2}{\mu^2}\biggr]^{D/2-2}\nonumber\\
=\frac{e^2}{2\pi^2}\biggl[\frac{1}{6}\frac{2}{4-D} - \frac{1}{6}\gamma -\frac{1}{6}\ln(2\pi)
-\int^1_0dxx(1-x)\ln\biggl[x(1-x)\frac{p^2}{\mu^2} + \frac{m_t^2}{\mu^2}\biggr]\nonumber\\
+ O(4-D)\biggr].
\label{Dimreg}
\end{align}
Here, the term $2/(4-D)$ in (\ref{Dimreg}) is divergent as $D\rightarrow 4$, whereas the logarithmic term $\ln(\Lambda^2_W/\mu^2)$ in
(\ref{regularized}) is finite for a measured value of $\Lambda_W$.
As in the usual EW calculations, we can remove the latter contribution as well as the other contributions not included in the integral by a modified
minimal subtraction $\overline{MS}$~\cite{Aitchison}.

A numerical integration of (\ref{gammavac}) demonstrates that $I(p^2)$ is positive for large $p^2$ and $I(p^2)\rightarrow 0$ as $p^2\rightarrow\infty$. We get from (\ref{runningconst}) for $p^2\rightarrow\infty$:
\begin{equation}
\label{alpha}
\alpha(p^2)\rightarrow\alpha_0.
\end{equation}
We observe that we do not have a Landau pole in our model. This can be important for our embedding of local QED in the larger group $SU_c(3)\times SU(2)_L\times U_{\rm em}(1)$, because it avoids a triviality problem. When an electron comes close to a nucleus as $p^2\rightarrow\infty$, the charge of the nucleus is anti-screened {\it and our QED is asymptotically safe.} This is analogous to the anti-screening that occurs in QCD leading to asymptotic freedom. When the interactions with quarks and leptons and the gluon self-energy are taken into account, the strong QCD coupling constant $\alpha_s(p^2)$ will not have a Landau pole and the colored $SU_c(3)$ will be asymptotically safe~\cite{Woodard}.

Radiative corrections alter the $\rho$ parameter determining the relative strength of the charged to neutral currents $J^\mu_ZJ_{\mu Z}/J^{\mu
+}J_\mu^{-}$:
\begin{equation}
\rho=\rho_{(0)} + \Delta\rho_{(1)},
\end{equation}
where $\rho_{(0)}=1$ and $\Delta\rho_{(1)}$ denotes the one-loop correction dominated by the top quark. In the standard EW Higgs model there is an extra contribution coming from the Higgs particle:
\begin{equation}
\label{Higgsdelta}
\Delta\rho_{H(1)}\sim -\frac{3G_FM_W^2}{8\pi^2\sqrt{2}}\biggl[\biggl(\frac{M_Z^2}{M_W^2}-1\biggr)\ln\biggl(\frac{m_H^2}{M_W^2}\biggr)\biggr],
\end{equation}
where $m_H$ denotes the Higgs mass, $M_W=80.398$ GeV and  $M_Z=91.1876$ GeV~\cite{pdg}. Also, we have $M_Z^2/M_W^2-1=\sin^2\theta_w/\cos^2\theta_w$.

For a light Higgs mass, $m_H \sim 116 - 135\,{\rm GeV}$, the non-oblique radiative Higgs corrections are not important. An example of this is
$Z\rightarrow b +{\bar b}$ decay. The Higgs loop corrections for this process for the decay of a neutral light Higgs are proportional to the coupling
$\lambda_b\sim \sqrt{2}m_b/v$ where $v=246$ GeV and are negligible~\cite{Logan}. Therefore, there is no need for these non-oblique radiative Higgs
corrections and they can be omitted in our minimal Higgless model. We shall concentrate on the oblique radiative corrections involving vacuum
polarization. The global fits to the low-energy EW data yield a light Higgs mass $m_H \sim 77 - 125$ GeV~\cite{EWworkgroups,Tevatron}. We see that for $m_H\sim M_W$ the Higgs contribution (\ref{Higgsdelta}) becomes negligible. For $m_H=125$ GeV, we obtain from (\ref{Higgsdelta}) $\Delta\rho_{H(1)}=-5.461\times 10^{-4}$. In our Higgless model the contribution (\ref{Higgsdelta}) is absent. We obtain in our model~\cite{Moff4}:
\begin{equation}
\rho\sim 1.01.
\end{equation}

\section{Unitarity of Scattering Amplitudes}

The standard EW model violates unitarity in scattering processes that involve longitudinally polarized vector bosons without the Higgs particle. The
scattering of two longitudinally polarized vector bosons $W_L$ results in a divergent term proportional to $s$. A less rapid divergence, proportional
to $\sqrt{s}$, occurs when fermions annihilate into a pair of $W_L$ vector bosons. The tree-level processes involving the Higgs boson in the standard
Higgs model cancel these divergences. A theory that does not incorporate a physical scalar Higgs particle must offer an alternative mechanism to either
cancel or suppress the badly behaved terms to maintain unitarity.

The scattering amplitude matrix elements for the process $W^+_L + W^-_L\rightarrow W^+_L + W^-_L$ is given in the SM by~\cite{MoffToth2}:
\begin{equation}
i{\cal M}_W=ig^2\left[\frac{\cos\theta+1}{8M_W^2}s+{\cal O}(1)\right],
\label{eq:MM2}
\end{equation}
where $\theta$ is the scattering angle. This result clearly violates unitarity for large $s$.
In the standard Higgs model, this behavior is corrected by the addition of the $s$-channel Higgs exchange in the high-energy limit:
\begin{equation}
i{\cal M}_H=-ig^2\left[\frac{\cos\theta+1}{8M_W^2}s+{\cal O}(1)\right].
\end{equation}
This cancels out the bad behavior in (\ref{eq:MM2}). In our EW model the unitarity violation is canceled by the high energy behavior of ${\bar g}(s)$.
The amplitude is now
\begin{equation}
\label{NewAmp}
i{\cal M}_W=i{\bar g}^2(s)\left[\frac{\cos\theta+1}{8M_W^2}s+{\cal O}(1)\right].
\end{equation}
We require that for $ \sqrt{s} > 1$ TeV, ${\bar g}(s)$ decreases as $\sim 1/\sqrt{s}$ or faster, resulting in the cancelation of the unitarity
violating contribution in Eq.(\ref{NewAmp}). We note that the $W$ mass $M_W$ in (\ref{NewAmp}) is the rest mass of an incoming $W$ boson with
3-momentum $|{\bf p}|=\sqrt{s/4-M_W^2}$ and it does not run with $s$.

We expect that a consistent choice of the entire function ${\cal E}(s/\Lambda_W^2)$ will lead to a different prediction for the $W^+_L +
W^-_L\rightarrow W^+_L + W^-_L$ scattering amplitude for $\sqrt{s} > 1$ TeV compared to the Higgs EW model, providing an experimental test of our
model.

\section{Conclusions}

By introducing generalized EW coupling constants ${\bar g}(p^2)$ and ${\bar g}'(p^2)$ which are energy dependent at Feynman diagram vertices with an
energy scale $\Lambda_W > 1$ TeV, we can obtain a Higgless  EW model that is unitary, finite and Poincar\'e invariant to all orders of perturbation
theory, provided that the coupling functions are composed of entire functions of $p^2$ that avoid any unphysical particles that will spoil the
unitarity of the scattering amplitudes. All the physical EW fields are local fields that satisfy microcausality. There is no Higgs particle in the
particle spectrum and this removes the troublesome aspects of the standard EW model with a spontaneous symmetry breaking. We do not attempt in this
version of the EW model to generate the $W$ and $Z$ boson masses or the quark and lepton masses. The measured masses of the particles, the measured
coupling constants $e$, $g$ and $g'$ and the energy scale $\Lambda_W$ are the basic constants of the model. This reduces the number of needed
parameters in the model compared to the standard EW model, for we do not postulate a Yukawa Lagrangian to generate quark and lepton masses with its
associated 12 coupling parameters. There are no anomalies in the model as is the case with the standard EW model with an equal number of quark and
lepton generations.

Because there is an absence of {\it physical} scalar fields in our model, only asymptotically safe, local gauge fields such as the the boson fields
$W,Z$ and $\gamma$ are present. In this way, we avoid certain pathologies associated with scalar fields and the QED photon field. Although the entire
function at the Feynman diagram vertices associated with the couplings ${\bar g}(p^2)$ and ${\bar g}'(p^2)$ is non-local, the vertex operators do not
describe propagating particles.  

We need to discover more information about the non-local entire function ${\cal E}(p^2)$. We must search for a way to obtain ${\cal E}(p^2)$ from some
underlying physical principle. An early attempt was made to obtain from a ``superspin" QFT entire functions that damp the Euclidean momentum loop
integrals~\cite{Moff5}. The knowledge of this function will determine the predicted scattering amplitudes and cross sections for such processes as
$W^+_L + W^-_L\rightarrow W^+_L + W^-_L$ and $e^+ + e^-\rightarrow W^+_L + W^-_L$ for $\sqrt{s} > 1$ TeV, which can be compared to the predictions of
the standard EW model with a Higgs particle. An important prediction made by our finite QFT is that if amplitudes and cross sections are experimentally consistent with a coupling constant ${\bar g}(s)$ that deviates from the behavior: ${\bar g}(s)\sim 1/\ln(s)$ for large $s$, then this will prove that
QFT does not conform to the standard renormalizable theory at high energies.

We have left as unknown the origin of fermion and boson particle masses. Because in our model there is no Higgs field pervading spacetime, then the
origin of particle masses must be sought in another physical phenomenon. Perhaps, gravity can be the source of particle masses. The gravitational and
inertial masses of a particle are  anticipated to play a fundamental role in discovering the origin of particle mass.

In the event that the LHC detects a Higgs particle, then the standard EW model can be vindicated. On the other hand, if it is excluded then we must
consider a significantly different EW model in which new fundamental properties of QFT will play a decisive role.

\section*{Acknowledgements}

I thank Viktor Toth, Martin Green, Roberto Percacci, Ruggero Ferrari and Robert Mann for helpful and stimulating discussions. This work was supported
by the Natural Sciences and Engineering Research Council of Canada. Research at the Perimeter Institute for Theoretical Physics is supported by the
Government of Canada through NSERC and by the Province of Ontario through the Ministry of Research and Innovation (MRI).

%\end{fmffile}


\begin{thebibliography}{100}

%\bibliography{refs}
%\bibliographystyle{unsrt}


\bibitem{LEP} R. Barate et al. [ALEPH, DELPHI, L3 and OPAL Collaborations, LEP Working Group for Higgs boson searches]. Phys. Lett. B {\bf 565} (2003)
    61 [arXiv:hep-ex/0306033].

\bibitem{EWworkgroups} ALEPH, CDF, D0, DELPHI, L3, OPAL and SLD Collaborations, LEP Electroweak Working Group, Tevatron Electroweak Working Group and
    LD Electroweak and Heavy Flavour Groups, arXiv:0911.2604[hep-ex].

\bibitem{Tevatron} [The TEVNPH Working Group of the CDF and D0 Collaborations], arXiv:1007.4587
[hep-ex]; B. Kilminster, ICHEP, Paris, France, July 2010.

\bibitem{Grojean} See, for example, C. Grojean, arXiv:0910.4976 [hep-ph] and references therein.

\bibitem{Moff} J. W. Moffat, Mod. Phys. Lett. A, {\bf 6}, 1011 (1991).

\bibitem{Moff2} J.W. Moffat, arXiv/0709.4269 [hep-ph].

\bibitem{MoffToth}  J. W. Moffat and V. T. Toth, arXiv:0812.1991 [hep-ph].

\bibitem{MoffToth2} J. W. Moffat and V. T. Toth, arXiv:0812.1994 [hep-ph].

\bibitem{Ferrari}  D. Bettinelli, R. Ferrari and A. Quadri, Phys. Rev. D{\bf 79}:125028 (2009), arXiv:0903.0281 [hep-th].

\bibitem{Weinberg} S. Weinberg, Phys. Rev. Lett. {\bf 19}, 1267 (1967);

\bibitem{Salam} A. Salam, {\it Elementary Particle Physics} ed. N. Svartholm (Stockholm: Almqvist and Wiksells) 1968.

\bibitem{Halzen} F. Halzen and A. D. Martin, {\it Quarks and Leptons: An introductory course in Modern Particle Physics}, John Wiley \& Sons, New York, 1984.

\bibitem{Aitchison} I. J. R. Aitchison and A. J. G. Hey, {\it Gauge theories in Particle Physics, Volume II: QCD and the Electroweak Theory, 3ed.},
Taylor and Francis, UK, 2004.

\bibitem{QuantGrav} J. W. Moffat, Eur. Phys. J. Plus 126:43 (2011), arXiv:1008.2482 [gr-qc].

\bibitem{Burnel} A. Burnel, Phys Rev. {\bf D33}, 2985 (1986). The explicit breaking of the $SU(2)_L\times U(1)_Y$ gauge
    symmetry by the mass contributions can be avoided by introducing a scalar field $\Phi$ into the Lagrangian and restoring gauge invariance. However, for the non-Abelian $SU(2)\times U(1)$ sector, this would result in scalar field $\Phi$ interactions and a possible Higgs particle in the particle
    spectrum. We avoid this possible alternative scenario and restrict ourselves to the Proca field quantization and the dynamical symmetry breaking
    scheme invoked though the vector boson masses.

\bibitem{thooft} G. 't Hooft, Nucl. Phys. {\bf B33}, 173 (1971).

\bibitem{thooft2} G. 't Hooft, Nucl. Phys. {\bf B35}, 167 (1971).

\bibitem{thooft3} G. 't Hooft and M. Veltman, Nucl. Phys. {\bf B44}, 189 (1972).

\bibitem{thooft4} G. 't Hooft and M. Veltman, Nucl. Phys. {\bf B50}, 318 (1972).

\bibitem{Titchmarsh} E. Titchmarsh, {\it Theory of Functions}, 2nd edition, Oxford University Press, 1939.

\bibitem{Boas} R. P. Boas, Jr. {\it Entire Functions} Academic Press Inc., New York, N.Y. 1954.

\bibitem{Holland} A. S. B. Holland, {\it Introduction to the Theory of Entire Functions}, Academic Press, New York and London, 1973.

\bibitem{Moff3} J. W. Moffat, Phys. Rev. {\bf D41}, 1177 (1990).

\bibitem{MoffWoodard} D. Evens, J. W. Moffat, G. Kleppe and R. P. Woodard, Phys. Rev. {\bf D43}, 499 (1991).

\bibitem{Cutkosky} R. E. Cutkosky, J. Math. Phys. {\bf 1}, 429 (1960).

\bibitem{Greiner} W. Greiner and J. Reinhardt, {\it Field Quantization}, (Springer) 1996.

\bibitem{Dirac} P. A. M. Dirac, {\it Lectures on Quantum Mechanics}, (Dover) 2001.

\bibitem{Jaffe} A. M. Jaffe, Phys. Rev. Lett. {\bf 17}, 661 (1966).

\bibitem{Jaffe2} A. M. Jaffe, Phys. Rev. {\bf 158}, 1454 (1967).

\bibitem{Meiman} N. N. Meiman, Zh. Eksp. Teor. Fiz. {\bf 47}, 1966 (1964).

\bibitem{Efimov} G. V. Efimov, Commun. Math. Phys. {\bf 5}, 42 (1967).

\bibitem{Efimov2} G. V. Efimov, Commun. Math. Phys.{\bf 7}, 138 (1968).

\bibitem{Efimov3} G. V. Efimov, Ann. Phys. (N.Y.) {\bf 71}, 466 (1972).

\bibitem{Hieu} N. V. Hieu, Ann. Phys. (N.Y.) {\bf 33}, 428 (1965).

\bibitem{Taylor} J. G. Taylor, Ann. Phys. (N.Y.) {\bf 68}, 484 (1971).

\bibitem{Constantinescu} F. Constantinescu, J. Math. Phys. {\bf 12}, 293 (1971).

\bibitem{Taylor2} J. G. Taylor and F. Constantinescu, Commun. Math. Phys. {\bf 30}, 211 (1973).

\bibitem{pdg} Phys. Lett. {\bf B667}, 1 (2008) [http:// pdg.lbl.gov].

\bibitem{Woodard} G. Kleppe and R. P. Woodard, Nucl. Phys. {\bf B388}, 81 (1992), arXiv:hep-th/9203016.

\bibitem{Logan} H. Haber and E. Logan, Phys. Rev. {\bf D62}, 015011 (2000).

\bibitem{Moff4} J. W. Moffat, arXiv:1103.0979 [hep-ph].

\bibitem{Moff5} J. W. Moffat, Phys. Rev. {\bf D39}, 3654 (1989).


\end{thebibliography}
\end{document}